\documentclass[twoside]{ilcws08}
\usepackage[latin1]{inputenc}
\usepackage[dvips]{graphicx,epsfig,color}
\usepackage{wrapfig,rotating}
\usepackage{amssymb,amsmath,array}

\begin{document}

\def\ti{\tilde}
\def\nt{\tilde \chi^0}
\def\ch{\tilde \chi^+}
\def\stau{\tilde\tau}
\def\cx    {\ti {\chi}} 
\def\tstau {\theta_{\ti\tau}} 
\def \Eslash {E^{\rm miss}}  
\def\ee{e^+e^-} 
\newcommand{\mch}[1]   {m_{\tilde\chi^\pm_{#1}}}
\newcommand{\cP } {{\cal P}} 
\newcommand{\GeV} {\mathrm{GeV}}

%\begin{flushright}
%  DESY ..........\\
%  LAPTH-Conf-1306/09\\
%  LPSC 08212\\
%\end{flushright}

\title{
%%%%   Paper title goes here  %%%%%%%%%%%%%%
Neutralino Relic Density in the CPVMSSM at the ILC} %% 
%***********************************************************************
% AUTHORS INFORMATION AREA
%***********************************************************************
\author{G.~B\'elanger$^1$, O.~Kittel$^2$, S.~Kraml$^3$, H.~U.~Martyn$^4$ and A.~Pukhov$^5$ 
% Optional short acknowledgment: remove next line if non-needed
% DO NOT MODIFY THE FOLLOWING '\vspace' ARGUMENT
\vspace{.3cm}\\
% Addresses and institutions (remove "1- " in case of a single institution)
1- LAPTH, Univ. de Savoie, CNRS \\
B.P. 110, F-74941 Annecy-le-vieux - France
%% Remove the next three lines in case of a single institution
\vspace{.1cm}\\
2- Departamento de F\'isica Te\'orica y del Cosmos and CAFPE,\\ 
Universidad de Granada, E-18071 Granada, Spain 
\vspace{.1cm}\\
3- LPSC, Univ. Joseph-Fourier,CNRS/IN2P3, INPG  \\
53 avenue des Martyrs, F-38026 Grenoble - France
\vspace{.1cm}\\
4- I. Physikalisches Institut, RWTH Aachen  \\
Sommerfeldstrasse 14, D-52074 Aachen - Germany
\vspace{.1cm}\\
5- SINP, Moscow State Univ.  \\
Moscow 119992 - Russia\\
}
%%***********************************************************************
% END OF AUTHORS INFORMATION AREA
%***********************************************************************

\maketitle

\begin{abstract}
We discuss ILC measurements for a specific MSSM scenario with CP phases, 
where the lightest neutralino, a candidate for dark matter, annihilates  
through t-channel exchange of light staus. These prospective  ILC 
measurements are  used to fit the underlying model parameters. A collider prediction 
of the relic density of the neutralino from this fit gives $0.116<\Omega h^2<0.19$ at 95\% CL. 
\end{abstract}

\section{Introduction}

One of the prime motivations for a  high-luminosity
$e^+e^-$ linear collider is the possibility to do precision
measurements of new particles beyond the Standard Model and in particular of the dark matter (DM) candidate. 
A precise determination of 
the properties of the DM in the laboratory could be used to
make a ``collider prediction" of its relic abundance,
which can be tested against cosmological models. For such a
collider prediction to be of interest, it must be at least as precise
as the value obtained from cosmological observations,  about 10\% 
for WMAP+SDSS~\cite{Spergel:2006hy}  and
a few percent  at the PLANCK satellite \cite{planck:2006uk}.

The possibility to make collider predictions of the cross sections
for annihilation of dark matter candidates has been examined
within specific supersymmetric scenarios. For example,
in  the general MSSM, it was shown that in a favourable
scenario the LHC could match roughly the WMAP+SDSS precision~\cite{Nojiri:2005ph} 
while the ILC could achieve much better precision~\cite{Baltz:2006fm}. 
These conclusions, however, depend
very strongly on the scenario considered; many remain challenging
even at the ILC~\cite{Baltz:2006fm}.

These studies  assumed that CP is conserved, although
CP-violating (CPV) phases are generic in the MSSM. 
%and are needed to generate the baryon asymmetry in the universe.
CPV MSSM phases can have an important effect on the neutralino annihilation cross
sections~\cite{Falk:1995fk} and could lead to  variations in
$\Omega h^2$ solely from modifications in the couplings of up to
an order of magnitude~\cite{Belanger:2006qa}. Therefore the determination
of the relevant couplings (including phases) can be as important for the prediction of $\Omega h^2$ as 
measurements of masses. In a particular scenario of the
CPV MSSM, the precision to which the underlying MSSM parameters could be determined from measurements at the ILC
was investigated in~\cite{Belanger:2008yc}. The resulting presicion on the neutralino relic density that could be inferred from
this was examined. The main reults are summarised below.

%--------------------------------------------------------------------------------------------------------------
\section{The stau-bulk scenario of the CPVMSSM}
\label{sec:benchmark}
%--------------------------------------------------------------------------------------------------------------

The scenario we investigate belongs to the ``stau bulk'' region of \cite{Belanger:2006qa},
which appears for light staus and large phase of $M_1$, the gaugino mass. 
The neutralino LSP is dominantly bino and annihilates 
predominantly into tau pairs, via t-channel exchange of
both $\stau_1$ and $\stau_2$. The annihilation cross section
is sensitive to the stau mixing .
%is crucial in bringing $\Omega h^2$ to the
%desired range, $0.094 < \Omega h^2 < 0.136$~\cite{Hamann:2006pf}.
Although the $\stau_1$ is very light, the scenario does not rely on
coannihilation (the mass difference with the LSP is too large). 

As a benchmark point  we choose the following
input parameters at the electroweak scale:
\begin{equation}
\begin{array}{lllll}
  M_1 = 80.47~{\rm GeV},\quad  & M_2 = 170.35~{\rm GeV},\quad & M_3 = 700~{\rm GeV},\quad
  &  \phi_1 = \pi, \quad & \mu = 600~{\rm GeV} \\
  M_{\tilde L_3} = 138.7~{\rm GeV},  & M_{\tilde E_3} = 135.2~{\rm GeV},  & A_\tau= 60~{\rm GeV},
  & \phi_\tau = 0, &  \tan\beta = 10.
%  M_{\tilde Q_3} = M_{\tilde U_3} = M_{\tilde D_3} & = A_{t,b} = 1~{\rm TeV}.
\end{array}
\label{eq:defpar}
\end{equation}
Other sfermion masses, $A_{t,b}$, and $m_{H^+}$ are set to 1~TeV  and
$\phi_\mu=0$.   
This way EDM constraints are avoided when varying $\phi_1$ and $\phi_\tau$.  
%The resulting mass pattern, light staus but TeV-scale selectrons/smuons,
%is not found in SUSY models where universality among scalar masses is imposed
%at a high scale. 
The particles accessible at ILC500  are $\nt_1$, $\nt_2,\ch_1$, $\stau_1$,
$\stau_2$ and $\tilde{\nu}_\tau$ with masses respectively of 
(80.7,\,164.9,\,164.9,\,100.9,\,177.2,\,123.1)~GeV as computed with {\tt CPsuperH}~\cite{Lee:2003nt}.  
Despite having a light spectrum this scenario is quite challenging for colliders.
Production of sparticles at ILC500 only lead to $\tau$'s  plus $E_T^{\rm miss}$ and similarly for
the cascade decays of squarks at LHC.

The relic density of the $\nt_1$ computed with 
{\tt micrOMEGAs2.2}~\cite{Belanger:2006is} is $\Omega h^2=0.130$.
The precision with which $\Omega h^2$ can be inferred from ILC measurements 
depends not only on the accuracy of the sparticle spectroscopy 
but also on the determination of all parameters of the neutralino sector
($M_1$, $M_2$, $\mu$, $\tan\beta$, $\phi_1$) and of the stau
sector ($M_{\tilde\tau_L}$, $M_{\tilde\tau_R}$, $A_\tau$, $\phi_\tau$).
The dependence on $\phi_1$, and to a much lesser extent  $\phi_\tau$, 
originates from the $\nt_1\tilde{\tau}_{1,2}^{}\tau$ couplings.
In addition, particles which are too heavy to be produced at ILC could
have some influence. We assume here that the mass scale of the squarks and gluino  is known from
LHC and that the ILC excludes selectrons/smuons and heavy Higgs states
up the the kinematic limit. This leaves an overall uncertainty 
 from the unknown part of the spectrum of  $\delta\Omega /\Omega\simeq 7\%$.

\section{ILC measurements} 

\begin{wraptable}{l}{0.5\columnwidth} 
\centerline{\begin{tabular}{|l|l|} 
\hline 
channel & observables \\ 
\hline 
      $\stau_1^+\stau_1^-$ & $m_{\stau_1} = 100.92 \pm 0.40~\GeV$ \\
                                               & $m_{\nt_1} = 80.67 \pm 0.35~\GeV$ \\ 
                                           & $\cos2\,\tstau = -0.065 \pm 0.028 $ \\
                                               &$\cP_\tau = 0.64 \pm 0.035 $  \\ 
      $\stau_2^+\stau_2^-$ & $m_{\stau_2} = 176.9 \pm 9.1~\GeV$  \\ 
      $\cx^+_1\cx^-_1$ & $m_{\cx^\pm_1} = 164.88 \pm 0.015~\GeV$  \\ 
\hline
\end{tabular}} 
\vspace*{-2mm}
\caption{Achievable precisions at ILC.} 
\label{tab:measurements} 
\end{wraptable} 

At the ILC, sparticle masses can be measured either with threshold scans or with the endpoint method. The challenge in this scenario is to disentangle the various sources that lead to the 
the $\tau\tau\Eslash$~ topology. Scanning downwards in energy
in steps of 10~GeV while using  different beam polarisations 
allows to detect thresholds:  the $\sigma_{RL}$ polarisation mode 
for  $\stau_i^+\stau_j^-$ pair production 
and the  $\sigma_{LR}$ mode for $\tilde\chi^+ \tilde\chi^-$ production. 
In addition the mixing angle in the stau sector, $\cos2\tstau$, can 
be determined from measurements of the polarized cross 
section while  the measurement of the $\tau$ polarisation 
$\cP_\tau$ in the decay  $\stau_1 \to \tau \nt_1$ gives  
additional information on the stau and neutralino mixings \cite{Nojiri:1994it}. 
It is in particular useful to constrain the gaugino--higgsino composition 
of the LSP. It was shown in ~\cite{Belanger:2008yc}  that masses,  
polarisation $\cP_{\tau}$ and mixing  $\tstau$ can be accurately determined even with moderate 
integrated luminosity, see Table~\ref{tab:measurements}. This analysis used 
simulations based on {\sc Simdet}~4.02~\cite{simdet},
beam polarisations $(\cP_{e^-},\cP_{e^+}) = (0.8, 0.6)$ and includes  QED radiation, beamstrahlung
and SM background to W pairs.  

%$\tau$ decays and polarisation are treated with {\sc Tauola}~\cite{Jadach:1993hs}. 
%The SM background includes $\ee \to W^+ W^-$, both the  
% $\ee\to \tau^+\tau^- (\gamma)$. 
%and $\gamma\gamma$ background are  negligible .

%---------------------------------------------------------------------------------------------------------
\section{DM properties: f\/it to ILC observables}
%---------------------------------------------------------------------------------------------------------

We performed a fit to the ILC observables listed in Table~\ref{tab:measurements} and used 
a Markov Chain Monte Carlo (MCMC) method to probe the 8-dimensional parameter space. 
The free parameters are
$M_1,\,\mu,\,\tan\beta,\, M_{\tilde L_3},\, M_{\tilde R_3},\, A_\tau,\, \phi_1,\,\phi_\tau$, while 
$M_2$ is computed from $\mu$ and $\tan\beta$ to match the
extremely well measured value of $\mch{1}$.
The total $\tau\tau$ SUSY cross section with polarized beams was also included in the fit.  
%For the benchmark point, 
Moreover, we have $\sigma_{\tau\tau}=3.220\pm 0.046$~pb at $\sqrt{s}=400$~GeV 
including the systematic and statistical uncertainties  for 10~fb$^{-1}$.

%The other SUSY parameters are fixed to the values specified in section~\ref{sec:benchmark}. 

\begin{wrapfigure}{r}{0.46\columnwidth}
\vspace{-0.5cm}
\centerline{\includegraphics[width=0.46\columnwidth]{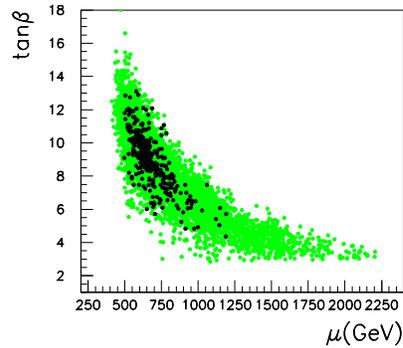}}
\vspace{-1cm}
\caption{$2\sigma$ (green) and $1\sigma$ (black) allowed region in the 
$\tan\beta$ vs $\mu$ plane. }\label{fig:tb}
\end{wrapfigure}

The very precise determination of $m_{\nt_1}$ and
$m_{\ch_1}$ constrain $M_1$ to $\delta
M_1=-0.9,+2.1$~GeV and indicates a
correlation between $\phi_1$ and $M_1$ although the phase $\phi_1$ can take any value. The measurement of the stau masses  and their mixing angle
constrain the stau soft masses  to about 10~GeV.
 The trilinear coupling $A_\tau$ and its phase $\phi_\tau$ are on the other hand basically undetermined. 
 This is because  the stau mixing which is proportional to 
$(A_\tau-\mu\tan\beta)$ is dominated by the term $\mu\tan\beta$. 
 This also means that $\mu$ and $\tan\beta$ are individually  poorly determined, see Fig.~\ref{fig:tb}.

\begin{figure}[b!]
\vspace{-0.1cm}
\centerline{\includegraphics[width=0.9\columnwidth]{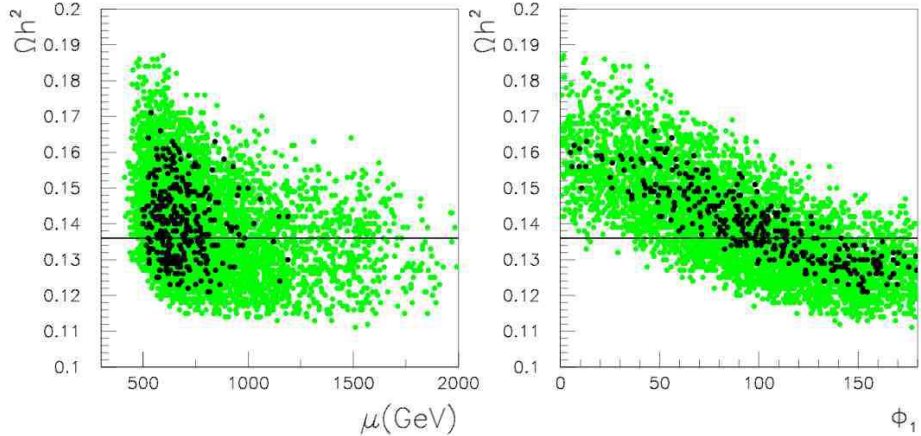}}
\vspace{-0.5cm}
\caption{Predictions for $\Omega h^2$ vs a) $\mu$ and b) $\phi_1$ and  the WMAP upper limit (line).}
\label{fig:omega}
\end{figure}

Within the allowed  parameter space, the relic density of dark matter is predicted to be 
$0.116<\Omega h^2<0.19$ at $2\sigma$. 
The largest values of  $\Omega h^2$ are found for $\phi_1\approx 0$ and small $\mu$, 
where the LSP has an increased higgsino component, see Fig.~\ref{fig:omega}.  
An improvement on the limit on the neutralino--proton cross section,  which is largest when $\mu$ is small,  
would cut on these scenarios thus somewhat reducing the range for $\Omega h^2$.

A better determination of $\tan\beta$ and $\mu$ would be needed to significantly improve the uncertainty in $\Omega h^2$. 
For both, a multi-TeV $e^+e^-$ linear collider offers the best prospects:
The parameter $\mu$ could be determined from a measurement of the heavy higgsino states. 
%A  measurement of the stop masses at the LHC and of the stop mixing angle at the ILC would completely determine the stop sector thus reducing the large parametric uncertainty in $m_h$.   
A  measurement of the stop masses and mixing angle at LHC and a multi-TeV LC would completely determine the stop sector thus reducing the large parametric uncertainty in $m_h$.   
The precise determination of $m_h$ at the ILC could then be used to constrain
$\tan\beta$.

CP-odd observables such as electric dipole moments or T-odd asymmetries could give a very clear signal of CP violation.
However, it was shown in~\cite{Belanger:2008yc} that a T-odd asymmetry based on triple-products in the production and decay of 
neutralinos~\cite{Kizukuri:1990iy} while sensitive to $\phi_1$,
would  not help constrain  $\Omega h^2$. The reason is a two-fold  ambiguity in $\phi_1$.

\section{Conclusions}

A precise determination of SUSY particle properties in the ``stau-bulk" scenario of the CPVMSSM
is quite challenging at the ILC.  Although  some of the underlying Lagrangian parameters 
can be extracted with very good precision, the large uncertainties in $\mu$, $\tan\beta$ and the 
phase $\phi_1$ induce a large uncertainty in the prediction of the neutralino relic density. 
To reduce this uncertainty to the percent-level would require 
precision  measurements of the heavy higgsino-like neutralinos and charginos 
at TeV energies.

\section{Acknowledgments}

This work was supported in part by GDRI-ACPP of CNRS,  
the ANR project ToolsDMColl,
the `SFB Transregio 33: The Dark Universe', and grants 
RFBR-08-02-00856-a, RFBR-08-02-92499-a of the Russian Foundation for Basic Research.

\begin{footnotesize}
% IF YOU DO NOT USE BIBTEX, USE THE FOLLOWING SAMPLE SCHEME FOR THE REFERENCES
% ----------------------------------------------------------------------------

\end{footnotesize}

% ****************************************************************************
% END OF BIBLIOGRAPHY AREA
% ****************************************************************************

\end{document}